\documentclass[runningheads]{llncs}

\usepackage{graphicx}
\usepackage{amsmath}
\usepackage{amssymb}
\usepackage{amsfonts}
\usepackage{hyperref}
\usepackage{comment}
\usepackage{tikz}

\newcommand{\da}{\Delta_\alpha}
\newcommand{\db}{\Delta_\beta}
\newcommand{\ra}{R_\alpha}
\newcommand{\rb}{R_\beta}

\newcommand{\eps}{\varepsilon}
\newcommand{\downto}{\downarrow}

\newcommand{\reals}{{\mbox{\bf R}}}

\newcommand{\Expect}{\mathop{\bf E{}}}

\newcommand{\eg}{{\it e.g.}}
\newcommand{\ie}{{\it i.e.}}

\newcommand{\BEAS}{\begin{eqnarray*}}
\newcommand{\EEAS}{\end{eqnarray*}}
\newcommand{\BEA}{\begin{eqnarray}}
\newcommand{\EEA}{\end{eqnarray}}
\newcommand{\BEQ}{\begin{equation}}
\newcommand{\EEQ}{\end{equation}}
\newcommand{\BIT}{\begin{itemize}}
\newcommand{\EIT}{\end{itemize}}

\begin{document}

 \title{Optimal Fees for Geometric Mean Market Makers}
\author{Alex Evans\inst{1}
\and
Guillermo Angeris\inst{2}
\and
Tarun Chitra\inst{3}}

\date{January 2021}

\institute{\email{alex@placeholder.vc}  \and
\email{guillean@stanford.edu } \and
\email{tarun@gauntlet.network}}

\maketitle

\begin{abstract}
Constant Function Market Makers (CFMMs) are a family of automated market makers that enable censorship-resistant decentralized exchange on public blockchains. Arbitrage trades have been shown to align the prices reported by CFMMs with those of external markets. These trades impose costs on Liquidity Providers (LPs) who supply reserves to CFMMs. Trading fees have been proposed as a mechanism for compensating LPs for arbitrage losses. However, large fees reduce the accuracy of the prices reported by CFMMs and can cause reserves to deviate from desirable asset compositions. CFMM designers are therefore faced with the problem of how to optimally select fees to attract liquidity. We develop a framework for determining the value to LPs of supplying liquidity to a CFMM with fees when the underlying process follows a general diffusion. Focusing on a popular class of CFMMs which we call Geometric Mean Market Makers (G3Ms), our approach also allows one to select optimal fees for maximizing LP value. We illustrate our methodology by showing that an LP with mean-variance utility will prefer a G3M over all alternative trading strategies as fees approach zero.

\end{abstract}

\section*{Introduction}

Constant Function Market Makers (CFMMs)~\cite{AC20} are a family of automated market makers that enable censorship-resistant decentralized exchange on public blockchains.
In CFMMs, Liquidity Providers (LPs) supply assets (reserves) to an on-chain smart contract. The smart contract makes reserves available for swaps, executing a trade only if it preserves some function of reserves, known as `the invariant.'
For example, Uniswap~\cite{uniswap} only permits trades that preserve the product of reserves (the product of reserve quantities must be the same before and after a trade). Similarly, Balancer only permits trades that preserve the weighted geometric mean of reserves. 
LPs are entitled to a pro-rata share of the CFMM's reserves, as well as any trading fees that the CFMM collects.
As of this writing, CFMMs have attracted billions of dollars worth of reserves and trade over \$1 Billion worth of cryptocurrency daily~\cite{dune_dexs}.
The rapid growth in the value deposited in CFMMs has allowed these protocols to regularly compete with established centralized exchanges on the basis of liquidity~\cite{uni_coinbase_theblock}.
At the same time, this growth has raised questions of efficiency, including how to optimally utilize reserves and how to select parameters for attracting liquidity and trading volume.

Under fairly general conditions, the prices reported by CFMMs have been shown to closely track those of external, more liquid markets~\cite{AC20}. Because the price reported by a CFMM is a function of reserves, this ``oracle" property requires an arbitrageur to maximize profit by adjusting reserves to align the CFMM's price with that of the external market. Because reserves are adjusted in response to price changes on the external market, the asset composition that Liquidity Providers (LPs) are entitled to is continually rebalanced. Protocols such as Balancer take advantage of this property to offer LPs payoffs that resemble constant-mix portfolios~\cite{balancer}. While LPs may benefit from rebalancing their portfolio to a target allocation, they also bear the cost of arbitrage transactions. In response, most CFMMs charge fees for incoming trades. However, fees make arbitrage less profitable, leading to only partial adjustment of reserves in response to price changes. This allows asset compositions to stray further from their desired allocations and reduces the accuracy of the prices reported by the CFMM. As a result, fees may in fact reduce the value LPs receive in certain cases. Given these trade-offs, CFMM designers are faced with the problem of how to assess the impact of fees on LP value and how to select optimal fees for attracting liquidity.

Our analysis focuses on Geometric Mean Market Makers (G3Ms), which include most popular CFMMs used in practice, including Uniswap, Sushiswap and Balancer \cite{uniswap,balancer}. G3Ms require that the reserves of the CFMM before and after each trade must have the same (weighted) geometric mean. 
As a by-product of arbitrage, the proportion of value deposited in the constant-mean CFMM for a given asset closely tracks the weight applied to the asset when calculating the weighted geometric mean~\cite{balancer}.
This property resembles a constant-mix portfolio and simplifies the analysis of Liquidity Provider (LP) returns. G3Ms allow us to model weight dynamics directly in the presence of fees. It has recently been shown that G3Ms can replicate a wide variety of trading strategies, including options payouts, using dynamic weights~\cite{evans2020liquidity}. Our analysis therefore extends naturally to a large class of LP payouts that can be represented by G3Ms.

\paragraph{Prior work.} Prior work on LP returns in CFMMs has primarily focused on the case where no fees are charged. In ~\cite{evans2020liquidity}, it is shown LPs in G3Ms with no fees underperform equivalent constant-mix portfolios due to arbitrage. 
However, the case with fees is  more involved as path independence is typically not satisfied. 
This question was addressed in~\cite{TW20} for the case of a Uniswap LP seeking to maximize the growth rate of wealth when the underlying price process follows a geometric Brownian motion.
This model assumes a particular functional form of the fee and shows that LPs can generate positive geometric growth with any non-zero fee provided that the mean and volatility are bounded in a suitable manner. The result holds for the specific case where the objective of the LP is to maximize the expected logarithm of reserve value when the underlying price process follows a geometric Brownian motion with certain mean and volatility constants. Our approach extends this setting to general diffusions and LP objective functions.

A separate line of work has applied conventional microstructure models to the problem of LP profitability, positing a game between LPs and informed traders to estimate profitability conditions for Uniswap~\cite{ja20}.
This framework is generalized to arbitrary CFMMs in~\cite{AEC20}, where it is shown that the curvature of the CFMM's trading function can be used to bound LP profitability. Our results apply to the case where the trader has perfect information and extracts risk-free profit at the expense of the LP. We show that there are general conditions under which the effect of arbitrage on the LP's value function approaches zero for small fees.

\paragraph{Optimal Control and Portfolio Optimization.}
While constant-mix portfolios produce excess growth due to rebalancing, no-fee G3Ms have been shown to divert this growth to arbitrageurs in order to incentivize continual rebalancing~\cite{evans2020liquidity}. While arbitrage losses are limited by increasing the fee that the G3M charges, this also limits the amount of rebalancing that arbitrageurs are incentivized to perform. We seek to formalize the impact of this trade-off on the value LPs receive from the G3M.

The problem of optimal portfolio selection in continuous time is well-studied in financial optimization, starting with the classical investment-consumption of Merton~\cite{Merton_1969}.
There are numerous extensions to the classical model that incorporate the impact of proportional transaction costs~\cite{david_norman_1990,Dumas_Luciano_1991,leland1999portfolio}.
In this setting, it is shown that the optimal investment policy involves a no-trade region around the optimal portfolio weight \cite{DIXIT1991657,dixit1993art,dumas91}.
In a G3M, an LP does not have direct control over reserves and relies on an arbitrageur for rebalancing reserves to the target asset proportions.
It is shown in~\cite{ACC19} that the arbitrageur takes no action inside a no-arbitrage interval around the desired weight.
However, the cost to the LP at the boundary is not proportional to the dollar value of rebalancing required.


The passive nature of LP rebalancing and non-proportional costs complicate the problem of optimal portfolio strategies for G3Ms.
Classical rebalancing~\cite{leland1999portfolio} assumes that the portfolio holder actively trades to adjust their portfolio weights.
Unlike traditional portfolio optimization, G3M arbitrageurs adjust the portfolio with the aim of extracting a profit at the expense of LPs.
In this work, we provide a solution to this problem by explicitly modeling the arbitrage costs incurred at the boundary of the no-trade region for different levels of the fee. 
Our approach is inspired by the stochastic control problems used in traditional portfolio optimization.
These methods are often used in reinforcement learning, portfolio analysis, and recently in decentralized finance (DeFi)~\cite{kao2020analysis}.

\paragraph{Summary.} We study the value to LPs of contributing capital to a G3M with fees. We consider the dynamics of the portfolio proportions as a function of time and fees assuming arbitrageurs trade against the CFMM to maximize profit. We show that the proportion of G3M value held in a given asset fluctuates freely within an interval where arbitrage is unprofitable. If the state variable exits this interval, an arbitrage opportunity arises to return it a point in the interior of the interval. We explicitly calculate the cost of this adjustment and show that it vanishes to first order when the state process has contiunous sample paths. The observation allows one to compute the value to the LP for a given choice of fee by solving a differential equation subject to two conditions that hold at the boundary of the no-arbitrage interval. We illustrate this approach for the specific example of maximizing mean-variance utility for a geometric Brownian motion and demonstrate how to optimize the resulting value.

\section{Problem description}\label{sec:problem}

In the arbitrage game, we have two players, each of whom trades in a two-asset economy: the liquidity provider, who is interested in minimizing some penalty function depending on the portfolio weights and an arbitrageur who trades against the liquidity provider's assets (and therefore, as a side effect, changes the portfolio weight).

\paragraph{Penalty function.} We will define the \emph{penalty function} $\phi: \reals \to \reals \cup \{+\infty\}$ which depends on the portfolio weight $w \in [0,1]$ for a given coin. The penalty function maps the weight to a liquidity provider's loss; \ie, we can view the function $\phi$ as the `tracking error' common in the control literature. We will assume that $\phi$ has a minimizer $w^\star \in [0, 1]$ such that $\phi(w^\star) \le \phi(w)$ for all $w \in [0, 1]$.

\paragraph{Portfolio weight dynamics.} The portfolio weight is, in general, a stochastic process that evolves in time, which we will write as $w_t \in [0, 1]$ at time $t$. We will assume a discretization in time, with steps of size $h > 0$ and later recover continuous results by taking the appropriate limits, such that $t=0, h, 2h, \dots$. In this case, we will assume a basic model with increments given by
\[
\xi_{t}(w, h) = a(w)h + b(w)\eps_t\sqrt{h}, \quad t=0, h, \dots,
\]
where $\eps_t \sim \{\pm 1\}$ with equal probability. (For example, if $a(w) = 0$ and $b(w) = 1$ for any $w$, then as $h \downto 0$ we have that $\sum_{n=1}^{\tau/h} \xi_{nh}$ converges weakly to a standard Brownian motion over time $\tau$.) Then, the dynamics of the weights will be given by some function $F: \reals \times \reals \to \reals$:
\[
w_{t+h} = F(w_t, \xi_t(w_t, h)), \quad t=0, h, \dots,
\]
where $F$ is a function that models the arbitrage dynamics; \ie, the arbitrageur sees a change in the portfolio weight of $\xi_t$ and performs arbitrage which results in some new weight $w_{t+h}$. We will often abuse notation slightly by writing $\xi_t$ instead of $\xi_t(w_t, h)$ to improve readability.

As a side note, we will be very informal regarding different types of convergence in the presentation and will freely switch expectations, limits, and derivatives, along with assuming that all functions are `nice enough.' While we will mostly work with the discrete approximations, some limits taken at the end will require justification---a reader familiar with stochastic processes and basic analysis should be able to insert the corresponding theorems as necessary, but we will not discuss them further.

\paragraph{Arbitrage loss and total expected loss.} By definition, the arbitrageur is guaranteed nonnegative profit at every time $t$ by exploiting the change in portfolio weights from time $t$ to $t+h$. We can (conversely) view this as a penalty incurred by the LP which we will call the \emph{adjustment cost}, defined by a nonnegative function $C: \reals \times \reals \to \reals_+$. A simple interpretation for $C(w_t, \xi_t) \ge 0$ is that it is the cost at time $t$, incurred by the LP, for adjusting the would-be portfolio weights $w_t + \xi_t$ to some different weight $w_{t+h}$.

This lets us write the expected loss for a liquidity provider, starting at weight $w \in [0, 1]$:
\begin{equation}\label{eq:expected-loss}
J(w) = \Expect\left[\sum_{n=0}^\infty e^{-nhr}(\phi(w_{nh})h + C(w_{nh}, \xi_{nh})) \biggm| w_0 = w\right],
\end{equation}
where $w_{t+h} = F(w_t, \xi_t)$. Here, $r$ is the continuous discounting rate, such that $e^{-rt}$ is the amount discounted at time $t$.

\paragraph{A (tight) lower bound.} A simple lower bound on the expected loss $J$ comes from the fact that $\phi(w_{nh}) \ge \phi(w^\star)$, by definition of $\phi(w^\star)$, and $C(w_{nh}, \xi_{nh}) \ge 0$ by definition of the adjustment cost $C$, which implies that the expected loss is bounded from below by:
\[
J(w) \ge \sum_{n=0}^\infty e^{-nhr}\phi(w^\star)h.
\]
If $\phi$ is normalized such that $\phi(w^\star) = 0$ (this can be done without loss of generality by replacing $\phi(w)$ with $\phi(w) - \phi(w^\star)$), this simplifies to:
\[
J(w) \ge 0.
\]
The remainder of the paper shows that, in fact, this simple bound becomes asymptotically tight as the fees approach, but do not equal, zero. (We will see soon how such fees connect to the problem.) This would immediately imply that the liquidity provider's losses are minimized by reducing the fees as much as possible, while ensuring they are not zero.

\paragraph{No-fee interval.} In general, CFMMs have a \emph{no-fee interval} (which is a function of the fees) where no possible weight adjustment is
profitable for arbitrageurs~\cite{ACC19,AC20}. For most CFMMs, and, more specifically, for the G3Ms we study here, the no-fee interval $[w_D, w_U] \subseteq [0, 1]$ has nonempty interior when the fee is nonzero; \ie, $w_D < w_U$. This condition implies that, if the portfolio weight $w_t$ lies in the interior of the interval, any vanishingly small change will not be adjusted and incurs no losses. More formally, if $w_D < w_t < w_U$, then
\begin{equation}\label{eq:fee-interval}
w_{t+h} = F(w_t, \xi_t) = w_t + \xi_t \qquad \text{and} \qquad C(w_t, \xi_t) = 0,
\end{equation}
for all $h$ small enough, since $\xi_t \sim O(h^{1/2})$ by definition. We will show this is true for all G3Ms in~\S\ref{sec:fees}.

\paragraph{Differential equation limit.} While~\eqref{eq:expected-loss} is a complete description of our problem, it is in general not easy to
analyze directly. On the other hand, in a similar way to dynamic programming, we can write $J(w_t)$ in terms of the current cost at time $t$ plus
a discounted expectation of $J(w_{t+h})$ given $w_t$:
\[
J(w) = \phi(w)h + \Expect[C(w, \xi_t) \mid w_t = w] + e^{-rh} \Expect[J(w_{t+h}) \mid w_t = w].
\]
By rearranging and dividing both sides by $h$, we find that
\begin{multline}
\label{eq:no-limit}
e^{-rh}\Expect\left[\frac{J(w_{t+h}) - J(w)}{h} \biggm| w_t = w\right] + \phi(w) \\
+ \frac{\Expect[C(w, \xi_t)\mid w_t = w]}{h} - \frac{1 - e^{-rh}}{h}J(w) = 0.
\end{multline}
Note that, if $w$ lies in the interior of the no-arbitrage integral, $w_D < w < w_U$, then the limit as $h \downto 0$ implies that $C(w, \xi_t)/h = 0$, since $\xi_t \downto 0$ and $C(w, \xi_t) = 0$ for all $\xi_t$ small enough, from~\eqref{eq:fee-interval}. Similarly, since $w_{t+h} = w_t + \xi_t$ for all $\xi_t$ small enough, we have
\[
\lim_{h \downto 0}\Expect\left[\frac{J(w_{t+h}) - J(w)}{h} \biggm| w_t = w\right] = a(w)J'(w) + \frac{b(w)^2}{2}J''(w),
\]
which follows from Taylor expanding $J(w_{t+h}) = J(w + \xi_t)$ into its linear and quadratic terms, as $\xi_t$ is on the order of $h^{1/2}$, and taking the corresponding expectation. This means that the final limit of~\eqref{eq:no-limit} as $h \downto 0$,
is the following differential equation:
\begin{equation}\label{eq:recursion}
a(w) J'(w) + \frac{b^2(w)}{2}J''(w) + \phi(w) - rJ(w) = 0,
\end{equation}
whenever $w_D < w < w_U$. In order to solve this differential equation, we
will also need appropriate boundary conditions which will depend on the
specifics of the CFMM we are considering. In our case, we will show that
\begin{equation}\label{eq:boundary-conditions}
J'(w_D) = J'(w_U) = 0,
\end{equation}
is satisfied.

\section{Fees for G3Ms}\label{sec:fees}
In this section, we will provide a specific application of the framework provided in~\S\ref{sec:problem} in order to show that a G3M with appropriately chosen
weights will always have an optimal fee that is as small as possible without being zero.

\paragraph{Constant function market makers.} A constant function market maker is defined by its reserves
$\ra(t)$ of coin $\alpha$ and $\rb(t)$ of coin $\beta$ at time $t$. Traders can trade with the CFMM (and therefore liquidity providers' funds) by proposing a trade $\da \ge 0$ of coin $\alpha$
and $\db \le 0$ of coin $\beta$ to the CFMM. The trade is accepted if the CFMM's \emph{trading function} defined by $\psi: \reals_+\times \reals_+ \to \reals$ satisfies
\[
\psi(\ra(t)+ \gamma_2\da, \rb(t) + \db) = \psi(\ra(t), \rb(t)).
\]
(\ie, it is `kept constant.') Here $(1-\gamma)$ is the fee, which must satisfy $0 < \gamma \le 1$. If, instead, we wish to trade $\da \le 0$ for $\db \ge 0$, we would instead switch the fee
to the incoming coin $\beta$, \ie, the trade is accepted if
\[
\psi(\ra(t)+ \da, \rb(t) + \gamma_1\db) = \psi(\ra(t), \rb(t)).
\]
If the trade $(\da, \db)$ satisfies either equation, then the CFMM takes $\da$ from the trader (if $\da \ge 0$, otherwise it pays out $\da$) and pays out $\db \ge 0$ (as before), updating its reserves to
\[
\ra(t+h) = \ra(t) + \da\quad \text{and} \quad \rb(t+h) = \rb(t) + \db.
\]
For more information on CFMMs see, \eg,~\cite{AC20}.

In the special case of G3Ms, which is the case we consider in the remainder of the paper, we have the specific trading function:
\[
\psi(\ra, \rb) = \ra^{1-\theta}\rb^{\theta},
\]
where $0 < \theta < 1$ is called the weight parameter. We derive explicit formulas for $w_U, w_D$ and the adjustment costs in this case in Appendix \S\ref{app:arb-results}.

\paragraph{Portfolio value and weight.} The definition of the \emph{portfolio value} of liquidity providers for the CFMM is the total present market value of reserves. If asset $\beta$ has some market value $S(t)$ at time $t$, then the portfolio value is given by
\[
\ra(t) + S(t)\rb,
\]
and the \emph{portfolio weight} (of coin $\beta$) of the liquidity providers is defined as
\[
w_t = \frac{\rb(t) S(t)}{\ra(t) + \rb(t) S(t)}.
\]
In other words, $w_t$ is the total proportion of wealth allocated to asset $\beta$ with respect to the complete porfolio.

\paragraph{Price process.}
We will compute the optimal fees when the price of the risky asset follows
\begin{equation}
S(t+h)=S(t)\left[(\mu-r)h + \sigma \eps_t \sqrt{h}\right],
\end{equation}
where $\eps_t \sim \{\pm 1\}$ is uniform and $\mu$, $r$ and $\sigma$ are constants that represent the growth rate, discounting rate, and volatility, respectively. (We will later take $h \downto 0$ such that $S(t)$ converges to a geometric Brownian motion.)

When no adjustments occur by the arbitrageur, there is no trade performed and so $\ra(t)=\ra$ and $\rb(t)=\rb$ are constant from $t$ to $t+h$. So, the corresponding dynamics of $w_t$ can be derived in the limit of small $h$:
\begin{equation}\label{eq:asset-props}
 w_{t+h} - w_{t} = w_t(1-w_t)(\mu-r-w_t \sigma^2)h+w_t(1-w_t)\sigma \eps_t\sqrt{h} + O(h^{3/2}).
\end{equation}
Through a discrete approximation, we prove the boundary conditions \eqref{eq:boundary-conditions} for these weight dynamics in Appendix \S\ref{app:main_proof}.

\paragraph{Penalty function.}
We assume the penalty function is given by
\begin{equation} \label{eq:phi}
\phi(w_t)=\frac{1}{2}\lambda\sigma^2(w_t-w^{*})^2
\end{equation}
for some constant $w^{*}$. This functional form is used in~\cite{leland1999portfolio} and conforms with the assumption that LP has mean-variance preferences over rates of return to wealth with risk aversion parameter $\lambda$. In particular, we note that this expression generalizes the setting considered in~\cite{TW20} where one seeks to maximize the growth rate of LP wealth. To see this, note the expected logarithm of wealth satisfies 
\begin{equation*}
    \frac{1}{T}\Expect[\ln(W(T)/W(0)]=\frac{1}{T}\int_0^{T} (w_s(\mu-r)-\frac{1}{2}\sigma^2w_s)ds.
\end{equation*}
Through a standard procedure, this expectation can be shown to be maximized by fixing $w^{*}=\frac{\mu-r}{\sigma^2}$. Substituting this value when taking the difference between the growth rate at $w^{*}$ and $w$,
\begin{equation*}
    [(\mu-r)w^{*}- \frac{1}{2}\sigma^2(w^{*})^2]dt-[(\mu-r)w_t- \frac{1}{2} \sigma^2w_t^2]dt=\frac{1}{2}\sigma^2(w-w^{*})^2dt,
\end{equation*}
which~\eqref{eq:phi} for the special case of $\lambda=1$.

\paragraph{Approximation.} We consider the case when $w_t \approx w^{*}$ which will provide a close approximation for small fees. When this is true, we can approximate \eqref{eq:asset-props} by
\[
  w_{t+h} - w_t = w_tah+w_tbd\eps_t\sqrt{h} \label{eq:w_process},
\]
where
\[
  a=(1-w^{*})(\mu-r-w^{*} \sigma^2), \qquad b=(1-w^{*})\sigma.
\]
The equation in~\eqref{eq:recursion} simplifies to Euler-Cauchy form and has an explicit solution given in~\cite{leland1999portfolio},
\begin{equation} \label{eq:general_solution}
    J(w,\gamma_1,\gamma_2)=\frac{1}{2}\lambda\sigma^2\left[\frac{w^2}{r-2a-b^2}-\frac{2ww*}{r-a}+\frac{(w^{*})^2}{r}\right]+C_1w^{z_1}+C_2w^{z_2}.
\end{equation}
where 
\begin{align*}
    z_1=\frac{\frac{b^2}{2}-a+\sqrt{(a-\frac{b^2}{2})^2+2b^2r}}{b^2}, \qquad
    z_2=\frac{\frac{b^2}{2}-a-\sqrt{(a-\frac{b^2}{2})^2+2b^2r}}{b^2}
\end{align*}
\paragraph{Determining optimal values.} Note that \eqref{eq:recursion} and \eqref{eq:boundary-conditions} will hold for all values of $\gamma_1$ and $\gamma_2$. For the optimal values, we show in \S\ref{app:optimality} that
\begin{align} \label{eq:optimality}
  J_{11}\left(w_U,\gamma_1,\gamma_2 \right)=J_{11}\left(w_D,\gamma_1,\gamma_2\right)=0.
\end{align}
Using the conditions in~\eqref{eq:boundary-conditions} and \eqref{eq:optimality}, we will determine the values of the coefficients $C_1, C_2$ in \eqref{eq:general_solution} as well as the optimal values for $\gamma_1$ and $\gamma_2$. One can check that the (numerical) maxima happen when $\gamma_1$ and $\gamma_2$ both approach 1 (zero fee). However, the system of equations has no solution as $C_1$ and $C_2$ are undefined for $\gamma_1=\gamma_2=1$. Taking the limit as $(\gamma_1,\gamma_2) \to (1^{-},1^{-})$, one can show that $J(w,\gamma_1,\gamma_2)$ approaches zero, implying that no cost is incurred relative to the optimal strategy.

\begin{figure}
    \centering
    \includegraphics[width=.9\textwidth]{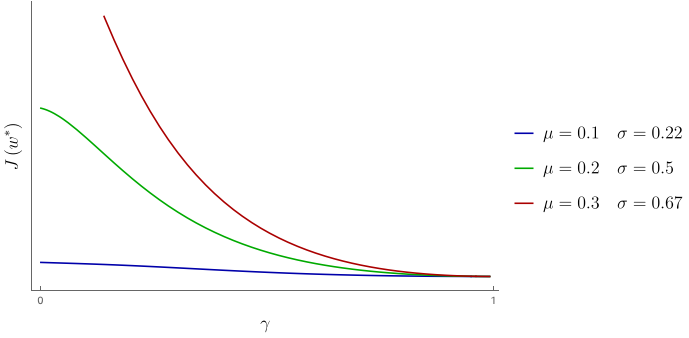}
    \caption{Cost function for $J(w^\star)$ when the LP seeks to minimize the penalty on the rate of wealth growth ($\lambda=1$) and fees are equal regardless of trading direction $\gamma=\gamma_1=\gamma_2$. We plot different mean-variance pairs that each satisfy $w^\star=\frac{1}{2}$. Higher volatility increases the relative cost incurred for higher values of the fee, for every choice $J(w^\star)$ is brought close to zero as the fee approaches zero (while the function is not continuous at $\gamma=1$)}
    \label{fig:J_plot}
\end{figure}

\section{Conclusion}
Fees are a critical component of LP value in CFMMs. Fees offset the cost of arbitrage, but also reduce the extent of rebalancing performed. We formalize this trade-off through a control-inspired approach that allows us to explicitly derive a solution for LP value for given fee choices. This solution also allows us to make the optimal choice of fees for maximizing value for the LP. In the example where the LP faces a quadratic tracking error for asset prices following geometric Brownian motion, we show that costs are minimized as fees approach zero. Our result applies to all G3Ms and allows one to derive results for general LP objective functions when the underlying asset price dynamics are governed by a continuous process.

\bibliography{references}

\begin{thebibliography}{10}
\providecommand{\url}[1]{\texttt{#1}}
\providecommand{\urlprefix}{URL }
\providecommand{\doi}[1]{https://doi.org/#1}

\bibitem{uniswap}
Adams, H.: Uniswap whitepaper. Tech. rep. (2018)

\bibitem{AC20}
Angeris, G., Chitra, T.: Improved price oracles: Constant function market
  makers. In: Proceedings of the 2nd ACM Conference on Advances in Financial
  Technologies. p. 80–91. AFT '20, Association for Computing Machinery, New
  York, NY, USA (2020). \doi{10.1145/3419614.3423251},
  \url{https://doi-org.stanford.idm.oclc.org/10.1145/3419614.3423251}

\bibitem{AEC20}
Angeris, G., Evans, A., Chitra, T.: When does the tail wag the dog? {C}urvature
  and market making. arXiv preprint arXiv:2012.08040  (2020)

\bibitem{ACC19}
Angeris, G., Kao, H.T., Chiang, R., Noyes, C., Chitra, T.: An analysis of
  {U}niswap markets. Cryptoeconomic Systems Journal p. to appear (2019)

\bibitem{ja20}
Aoyagi, J.: Lazy liquidity in automated market making. Available at SSRN
  3674178  (2020)

\bibitem{uni_coinbase_theblock}
Cermak, L.: Uniswap's monthly trade volume exceeded coinbase's in september.
  \url{https://www.theblockcrypto.com/linked/79775/uniswap-coinbase-monthly-volume-september}
  (October 2020)

\bibitem{david_norman_1990}
Davis, M.H.A., Norman, A.R.: Portfolio selection with transaction costs.
  Mathematics of Operations Research  \textbf{15}(4),  676--713 (1990),
  \url{https://EconPapers.repec.org/RePEc:inm:ormoor:v:15:y:1990:i:4:p:676-713}

\bibitem{dixit1993art}
Dixit, A.: The Art of Smooth Pasting. Fundamentals of pure and applied
  economics, Harwood Academic Publishers (1993),
  \url{https://books.google.com/books?id=nmoUDNfYK2sC}

\bibitem{DIXIT1991657}
Dixit, A.: A simplified treatment of the theory of optimal regulation of
  brownian motion. Journal of Economic Dynamics and Control  \textbf{15}(4),
  657 -- 673 (1991). \doi{https://doi.org/10.1016/0165-1889(91)90037-2},
  \url{http://www.sciencedirect.com/science/article/pii/0165188991900372}

\bibitem{dumas91}
Dumas, B.: Super contact and related optimality conditions. Journal of Economic
  Dynamics and Control  \textbf{15}(4),  675--685 (1991),
  \url{https://EconPapers.repec.org/RePEc:eee:dyncon:v:15:y:1991:i:4:p:675-685}

\bibitem{Dumas_Luciano_1991}
Dumas, B., Luciano, E.: An exact solution to a dynamic portfolio choice problem
  under transactions costs. Journal of Finance  \textbf{46}(2),  577--95
  (1991),
  \url{https://EconPapers.repec.org/RePEc:bla:jfinan:v:46:y:1991:i:2:p:577-95}

\bibitem{dune_dexs}
Dune: Dune analytics decentralized exchange dashboard (2020), available at
  \url{https://explore.duneanalytics.com/public/dashboards/c87JEtVi2GlyIZHQOR02NsfyJV48eaKEQSiKplJ7}

\bibitem{evans2020liquidity}
Evans, A.: Liquidity provider returns in geometric mean markets. arXiv preprint
  arXiv:2006.08806  (2020)

\bibitem{kao2020analysis}
Kao, H.T., Chitra, T.: Feedback control as a new primitive for {DeFi} (2020),
  \url{https://medium.com/gauntlet-networks/feedback-control-as-a-new-primitive-for-defi-27b493f25b1}

\bibitem{leland1999portfolio}
Leland, H.E.: Optimal portfolio management with transactions costs and capital
  gains taxes. Working Paper RPF-290, IBER, UC Berkeley, Available at SSRN:
  https://ssrn.com/abstract=206871  (December 1999)

\bibitem{balancer}
Martinelli, F., Mushegian, N.: Balancer: {{A}} non-custodial portfolio manager,
  liquidity provider, and price sensor  (2019)

\bibitem{Merton_1969}
Merton, R.: Lifetime portfolio selection under uncertainty: The continuous-time
  case. The Review of Economics and Statistics  \textbf{51}(3),  247--57
  (1969),
  \url{https://EconPapers.repec.org/RePEc:tpr:restat:v:51:y:1969:i:3:p:247-57}

\bibitem{TW20}
Tassy, M., White, D.: Growth rate of a liquidity provider's wealth in $xy = c$
  automated market makers.
  \url{https://math.dartmouth.edu/~mtassy/articles/AMM_returns.pdf} (Nov 2020)

\end{thebibliography}
\bibliographystyle{splncs04.bst}

\appendix

\section{G3M Arbitrage Results}\label{app:arb-results}
When the arbitrageur adds reserves of the risky asset, we have the following constant geometric mean formula,
\[
(\ra-\da)^{1-\theta}(\rb+\gamma_1\db)^{\theta}=\ra^{1-\theta}\rb^\theta.
\]
Solving for $\db$,
\[
\db=\frac{1}{\gamma_1}\rb\left( \left(\frac{\ra}{\ra-\da} \right)^{\frac{1-\theta}{\theta}}-1 \right)\]
The aribtrageur's problem is therefore
\begin{equation}\label{eq:balancer-arb}
\begin{aligned}
& \text{maximize} & &  \da - S(t)\frac{1}{\gamma_1}\rb\left(\left(\frac{\ra}{\ra-\da} \right)^{\frac{1-\theta}{\theta}}-1 \right)\\
& \text{subject to} & & \da \geq 0
\end{aligned}
\end{equation}
As in~\cite{ACC19}, we note that the unconstrained maxima are those where the derivative of \eqref{eq:balancer-arb} is zero. This happens when
\begin{equation}\label{eq:optimal-delta}
\da = \ra-\left(\frac{1-\theta}{\gamma_1\theta} S(t) \rb \ra^{\frac{1-\theta}{\theta}}    \right)^{\theta}.
\end{equation}
This implies
\[
\db = \left(\frac{\theta}{1-\theta}\frac{\ra}{S(t)}\right)^{1-\theta} \left(\frac{\rb}{\gamma_1}\right)^\theta-\frac{\rb}{\gamma_1}.
\]
Substituting this back into the objective of \eqref{eq:balancer-arb} and simplifying, we get that the total arbitrage profit for the trader
\[
\ra-\frac{1}{\theta^\theta(1-\theta)^{1-\theta}}\ra^{1-\theta}\left(\frac{S(t)\rb}{\gamma_1} \right)^\theta + \frac{S(t)\rb}{\gamma_1}
\]
Scaling to total LP wealth  $S(t)\rb + \ra$,
\begin{equation}\label{eq:down_cost}
C_d=(1-w(t))-\frac{1}{\theta^\theta(1-\theta)^{1-\theta}}\left(\frac{w(t)}{\gamma_1}\right)^\theta(1-w(t))^{1-\theta}+\frac{w(t)}{\gamma_1},
\end{equation}
where $w(t)=\frac{S(t)\rb}{S(t)\rb + \ra}$, the fraction of LP wealth in the risky asset prior to rebalancing. No-arbitrage requires that $\Delta \leq 0$ in \eqref{eq:optimal-delta}, which implies
\[
w(t) \geq \frac{\gamma_1 \theta}{1 - \theta + \gamma_1 \theta} =w_D
\]
After this adjustment, the quantities are updated to $\ra \mapsto \ra - \da$ and $\rb \mapsto \rb + \db$. The weight after the adjustment is given by
\begin{equation*}
    w_d(t) = \frac{(\rb+\db)S}{(\rb+\db)S+\ra-\da}
\end{equation*}
Which we can rewrite as
\begin{equation} \label{eq:adjusted-down}
    w_d(t) = \frac{1+\gamma_1^\theta \left(\frac{1-\theta}{\theta}\frac{w(t)}{1-w(t)} \right)^{1-\theta}(1-\gamma_1^{-1})} {\frac{1}{\theta}+\gamma_1^\theta \left(\frac{1-\theta}{\theta}\frac{w(t)}{1-w(t)} \right)^{1-\theta}(1-\gamma_1^{-1})}
\end{equation}
When adding units of the num\'eraire in exchange for the risky asset, the constant geometric mean gives
\[
(\ra+\gamma_2\da)^{1-\theta}(\rb-\db)^{\theta}=\ra^{1-\theta}\rb^\theta.
\]
Through a similar procedure, it is possible to show
\begin{equation}\label{eq:up_cost}
C_u=\frac{1}{\gamma_2}(1-w(t))-\frac{1}{\theta^\theta(1-\theta)^{1-\theta}}\left(\frac{1}{\gamma_2}(1-w(t))\right)^{1-\theta}w(t)^{\theta}+w(t),
\end{equation}
and
\[
w(t) \leq \frac{\theta}{\gamma_2 (1-\theta)+\theta}.
\]
The weight after the adjustment is given by 
\begin{equation*}
    w_u(t) = \frac{1}{\frac{1}{\theta}+\left(\gamma_2 \frac{1-\theta}{\theta}\right)^{1-\theta}\left( \frac{1-w(t)}{w(t)}\right)^{\theta}(1-\gamma_2^{-1})}.
\end{equation*}

\section{Proof of Boundary Conditions} \label{app:main_proof}
We proceed by a discrete approximation of the problem and derive the associated boundary conditions at the limit. Analogous to~\cite{DIXIT1991657,dixit1993art}, we divide time into discrete intervals of length $\tau$ and state into steps of size $\xi$. In what follows, we will approximate the weight variable in a slightly different way, but will still recover \eqref{eq:w_process} at the limit; here, for each $i$, we will have
\begin{equation*}
    w_{i+1}-w_i=\xi.
\end{equation*}
We approximate the unadjusted weight process with a random walk. Starting from $w_i$, the next step after $\tau$ units of time have passed will be $w_{i-1}$, with probability $p$, and $w_{i+1}$, with probability $q=1-p$. If we suppose these probabilities satisfy
\begin{align*}
    aw_i\tau = q \xi + p(- \xi),
\end{align*}
then this implies
\[
    p=\frac{1}{2}(1-aw_i \tau/\xi), \qquad q=\frac{1}{2}(1+aw_i \tau/\xi).
\]
The variance is given by
\[
    b^2w_i^2\tau = q(\xi-aw_i \tau)^2 + p(\xi+aw_i \tau)^2 = \xi^2 -a^2w_i^2 \tau^2.
\]
Keeping only the leading term, $b^2w_i^2\tau = \xi^2$,
and taking the limit as $\tau$ and $\xi$ tend to zero, we recover the process in \eqref{eq:w_process}. From \S\ref{app:arb-results}, we have the boundaries of the no-arbitrage interval,
\begin{align*}
\frac{\gamma_1 w^{*}}{1 - w^{*} + \gamma_1 w^{*}}=w_D, \qquad\frac{w^{*}}{\gamma_2 (1-w^{*})+w^{*}}=w_U.
\end{align*}
The random walk proceeds unadjusted on the states $i= D+1,.., U-1$. If the process is at $D$ and takes a step to the right, again no arbitrage adjustment occurs. If, however, the process moves to $D-1$, then arbitrage instantaneously adjusts the weight to $w_d$ in~\eqref{eq:adjusted-down}. At $i=D$ the next step will be $w_{D+1}$, with probability $p$, and $w_{d}$ with probability $q=1-p$. Similarly, at the upper boundary, we will have $w_{U-1}$ with probability $p$, and $w_{u}$, with probability $q=1-p$.
Therefore, at the boundary point $w_U$, we have
\begin{align*}
  J(w_U)=f(w_U)\tau+e^{-r\tau}pJ(w_{U-1})+e^{-r\tau}qJ(w_u(\xi))-qC_u(\xi),
\end{align*}
where
\begin{equation}\label{eq:cost_xi}
C_u(\xi)=\frac{1}{\gamma_2}(1-w_U-\xi)-\frac{(1-w_U-\xi)^{1-w^{*}}}{\gamma_2(w^{*})^{w^{*}}(1-w^{*})^{1-w^{*}}}(w_U+\xi)^{w^{*}}+w_U+\xi,
\end{equation}
and
\[
    w_u(\xi) = \frac{1}{\frac{1}{w^{*}}+\left(\gamma_2 \frac{1-w^{*}}{w^{*}}\right)^{1-w^{*}}\left( \frac{1-W_U-\xi}{W_U+\xi}\right)^{w^{*}}(1-\gamma_2^{-1})}.
\]
Rearranging terms and multiplying by $e^{r\tau}$
\[
  e^{r\tau}J(w_U)-pJ(w_{U-1})-qJ(w_u(\xi))=e^{r\tau} f(w_U)\tau - e^{r\tau}qC_u(\xi)).
\]
Expanding on the right side and noting that $\tau$ is $o(\xi)$,
\begin{multline*}
  \tau[1+r\tau+o(\tau)]f(w_U) -\frac{1}{2}(1+a \xi /(w_Ub^2))[1+r\tau+o(\tau)] C_u(\xi))\\
  = -\frac{1}{2}(1+a \xi /(w_Ub^2))C_u(\xi))+o(\xi).
\end{multline*}
Expanding on the left side,
\begin{align*}
  &[1+r\tau+o(\tau)]J(w_U)-p[J(w_U)-J_1(w_U)\xi+o(\xi)]+qJ(w_u(\xi))\\
  &= q[J(w_U)-J(w_u(\xi))]+pJ_1(w_U)+o(\xi) \\
  &= \frac{1}{2}(1+a \xi /(w_Ub^2))[J(w_U)-J(w_u(\xi))] + \frac{1}{2}(1-a \xi /(w_Ub^2))J_1(w_U)\xi+o(\xi)\\
  &=\frac12\left([J(w_U)-J(w_u(\xi))] + \frac{a \xi}{ w_Ub^2}[J(w_U)-J(w_u(\xi))] + J_1(w_U)\xi\right)+o(\xi).
\end{align*}
 Next, we divide both sides by $\xi$ and take the limit as $\xi$ tends to zero.  From \eqref{eq:cost_xi}, one can check that
\[
    C_U(\xi)=\frac{(\gamma_2(1-w^\star)-w)^3}{2\gamma_2^2(1-w^\star)w^\star}\xi^2 + o(\xi^2)
\]
Therefore, the right-hand side is zero. For the left-hand side, noting that $w_u(0)=w_U$ we have
\begin{align*}
  & \lim_{\xi \to 0} \frac12\left[ [J(w_U)-J(w_u(\xi))] + \frac{a \xi}{ w_Ub^2}[J(w_U)-J(w_u(\xi))] + J_1(w_U)\xi+o(\xi)\right] \\
  &= \lim_{\xi \to 0} \frac12\left[\frac{J(w_U)-J(w_u(\xi))}{\xi} + J_1(w_U) \right] 
\end{align*}
By the Mean Value Theorem, there exists $\zeta \in (w_U,w_u(\xi))$ such that
\begin{align*}
& \lim_{\xi \to 0} \frac12\left[\frac{J(w_U)-J(w_u(\xi))}{\xi} + J_1(w_U) \right] 
= \lim_{\xi \to 0} \frac12\left[\frac{J_1(\zeta)[w_U-w_u(\xi))]}{\xi} + J_1(w_U) \right] \\
&= \lim_{\xi \to 0} \frac12\left[-\frac{J_1(\zeta)[w_u'(0)\xi + o(\xi))]}{\xi} + \frac{1}{2}J_1(w_U) \right]
= \frac{1}{2}J_1(w_U)(1-w_u'(0)) \\
&=  \frac{\gamma_2(1-w^{*})-w^{*}}{2\gamma_2}J_1(w_U)
\end{align*} 
Noting $\frac{\gamma_2(1-w^{*})-w^{*}}{2\gamma_2}$ is non-zero and finite for $0 < \gamma_2 \leq 1$ completes the proof of \eqref{eq:boundary-conditions}. For $\gamma_2=0$, the boundary condition does not apply as no adjustment occurs and this holds for all $w>w_D=0$. The proof for the lower boundary is similar.

\section{Optimality conditions} \label{app:optimality}

Substituting the boundary condition \eqref{eq:boundary-conditions} into the general solution \eqref{eq:general_solution} we have

\[
    J_1(w_U,\gamma_1,\gamma_2)=\frac{1}{2}\lambda\sigma^2\left[\frac{2w_U}{r-2a-b^2}-\frac{2w^*}{r-a}\right]+C_1z_1w_U^{z_1-1}+C_2z_2w_U^{z_2-1}.
\]
Taking the derivative with respect to $\gamma_ 1$

\[
    J_{12}(w_U,\gamma_1,\gamma_2)=\frac{\partial C_1}{\partial \gamma_1}z_1w_U^{z_1-1}+\frac{\partial C_2}{\partial \gamma_1}z_2w_U^{z_2-1}=0.
\]
We note that for $b \neq 0$ and $r>0$, $z_1$ and $z_2$ will have opposite signs. Since $w^{z_1-q}$ and $w^{z_2-q}$ are positive, we conclude that $\frac{\partial C_1}{\partial \gamma_1}$ and $\frac{\partial C_2}{\partial \gamma_1}$ have the same sign. So,
\[
    J_{2}(w,\gamma_1,\gamma_2)=\frac{\partial C_1}{\partial \gamma_1}w^{z_1}+\frac{\partial C_2}{\partial \gamma_1}w^{z_2}.
\]
Again $w^{z_1}$ and $w^{z_2}$ are positive and the derivatives have the same sign. This implies that changing $\gamma_1$ either increases or decreases the total cost for all values of $w$. The first-order condition for optimality is therefore $J_{2}(w,\gamma_1,\gamma_2)=0$. We conclude that
\[
    \frac{\partial C_1}{\partial \gamma_2}=\frac{\partial C_2}{\partial \gamma_2}=0.
\]
Taking the derivative of \eqref{eq:boundary-conditions} with respect to $\gamma_1$ gives
\begin{align*}
 J_{12}(w_D,\gamma_1,\gamma_2)&=0 \\
 \frac{\partial C_1}{\partial \gamma_1}z_1w_D^{z_1-1}+\frac{\partial C_2}{\partial \gamma_1}z_2w_D^{z_2-1}+\qquad
 \\
 \frac{w^{*}(1-w^{*})}{(1+(\gamma_1-1)w^{*})^2}[C_1z_1(z_1-1)w_D^{z_1-2}2C_2z_2(z_2-1)w_D^{z_2-2}+\frac{ \lambda \sigma^2}{r-2a-b^2}]&=0 \\
 \frac{w^{*}(1-w^{*})}{(1+(\gamma_1-1)w^{*})^2}J_{11}(w_D,\gamma_1,\gamma_2)&=0,
\end{align*}
which gives the desired result for the second derivative at the lower boundary. The proof is identical for the upper boundary.


\end{document}